\documentclass[10pt, conference]{IEEEtran} % peerreview 10pt % conference
\IEEEoverridecommandlockouts
\usepackage{cite}
\usepackage{amsmath,amssymb,amsfonts}
\usepackage{algorithmic}
\usepackage{graphicx}
\usepackage{textcomp}
\usepackage{xcolor}
\usepackage{hyperref}
\usepackage{float}
\usepackage{enumitem}
\def\BibTeX{{\rm B\kern-.05em{\sc i\kern-.025em b}\kern-.08em
    T\kern-.1667em\lower.7ex\hbox{E}\kern-.125emX}}
\begin{document}

\title{A Static Analysis Platform for Investigating Security Trends in Repositories%\\
}

\makeatletter
\newcommand{\linebreakand}{%
\end{@IEEEauthorhalign}
\hfill\mbox{}\par
\mbox{}\hfill\begin{@IEEEauthorhalign}
}

\author{
\IEEEauthorblockN{Tim Sonnekalb\footnote{Corresponding author}, Christopher-Tobias Knaust,\\ Bernd Gruner, Clemens-Alexander Brust}
\IEEEauthorblockA{\textit{Institute of Data Science} \\
\textit{German Aerospace Center (DLR)}\\
Jena, Germany \\
\{firstname.lastname\}@dlr.de}
\and
\IEEEauthorblockN{Lynn von Kurnatowski, Andreas Schreiber}
\IEEEauthorblockA{\textit{Institute for Software Technology} \\
\textit{German Aerospace Center (DLR)}\\
Cologne, Germany \\
\{firstname.lastname\}@dlr.de}
%\and
\linebreakand
\IEEEauthorblockN{Thomas S. Heinze}
\IEEEauthorblockA{\textit{Cooperative University Gera-Eisenach}\\
Gera, Germany \\
thomas.heinze@dhge.de}
\and
\IEEEauthorblockN{Patrick M\"ader}
\IEEEauthorblockA{\textit{Data-intensive Systems and Visualization Group} \\
\textit{Technical University Ilmenau}\\
Ilmenau, Germany \\
patrick.maeder@tu-ilmenau.de}
}

\maketitle
%\IEEEpeerreviewmaketitle

\begin{abstract}
	Static analysis tools come in many forms and configurations, allowing them to handle various tasks in a (secure) development process: code style linting, bug/vulnerability detection, verification, etc., and adapt to the specific requirements of a software project, thus reducing the number of false positives.

	The wide range of configuration options poses a hurdle in their use for software developers, as the tools cannot be deployed out-of-the-box. However, static analysis tools only develop their full benefit if they are integrated into the software development workflow and used on regular. Vulnerability management should be integrated via version history to identify hotspots, for example.

	We present an analysis platform that integrates several static analysis tools that enable Git-based repositories to continuously monitor warnings across their version history. The framework is easily extensible with other tools and programming languages. We provide a visualization component in the form of a dashboard to display security trends and hotspots.
	Our tool can also be used to create a database of security alerts at a scale well-suited for machine learning applications such as bug or vulnerability detection.

\end{abstract}

\begin{IEEEkeywords}
static analysis, software security, automatic analysis, SAST tool database, security dashboard
\end{IEEEkeywords}

\section{Introduction}
Static Analysis Security Testing (SAST) tools provide powerful means for finding bugs and code flaws in written code. These tools have a wide range of features and range from code style linting to formal verification of the underlying code.
They allow for very detailed configuration settings and custom rules to account for the characteristics of a software project and maximize the detection rate. 

In contrast, few open source projects incorporate SAST tools into their development workflows, as found by a large-scale study \cite{beller2016analyzing}. The study also showed that when SAST tools were used, the default configuration was usually not adjusted. When custom rules were written, they correspond to only a small percentage of the total rule value. In addition, this study reveals that most configuration files are never adjusted again over the entire development period. In this way, static analysis cannot work effectively because the configuration does not match the requirements of the software.

Johnson et al. \cite{johnson2013don} found in interviews that developers find using these tools too burdensome and do not consider using bug tracking tools due to time constraints. Some would use such tools if they were already available in the development process and if the tool output is well-presented.

To dsvr developers most of the effort associated with using these code review tools, we develop an online static analysis platform that can be easily integrated into Git-based processes, run at regular intervals, and present the result via a dashboard.
In addition, our analysis platform can be used to mine SAST warnings at scale and leverage a comprehensive database for further data science applications such as empirical vulnerability detection.

We design a SAST analysis platform that satisfies the following aspects:
\begin{itemize}[itemindent=-1em] % , topsep=5pt, parsep=2pt
	\item[] \textbf{Continuous Online Monitoring} Our analysis checks for new commits for a given repository URL at regular intervals and analyzes them to support DevOps. 
	\item[] \textbf{High Coverage} Since we apply multiple SAST tools with different modes of operation, the probability of detecting a bug is higher, minimizing the number of false negatives \cite{Nunes2017}.
	\item[] \textbf{Virtualization} We deploy the analysis platform as Docker image to run it in an OS-independent manner.
	\item[] \textbf{Modular Design} Our implementation allows easy extension with various SAST tools and programming languages.
	\item[] \textbf{Git-based} We use Git as a widely used version control system and can apply our analysis to all Git-based repositories.
	\item[] \textbf{Mining/Visualization} The analysis results are written to a database and visualized in a dashboard.
\end{itemize}

Schreiber et al. \cite{SchreiberProvenance2021} tested this platform using the open source repositories of the German Covid contract tracing apps as case study. They also provided a database \cite{sonnekalb_tim_2021_5036046} which can be used with our dashboard.
Schreiber et al.  \cite{SchreiberDashboards2021} discussed elements for visualizing software components, including SAST warnings, which we used as the basis for our dashboard.
Our contribution in this work is the implementation and publication of the code for further work.
We provide the source code of our analysis platform, the dashboard and the Dockerfile via Gitlab \footnote{\url{https://gitlab.com/dlr-dw/sast-tool-platform}}.

\begin{figure*} %[h]
	\centering
	\includegraphics[width=\linewidth]{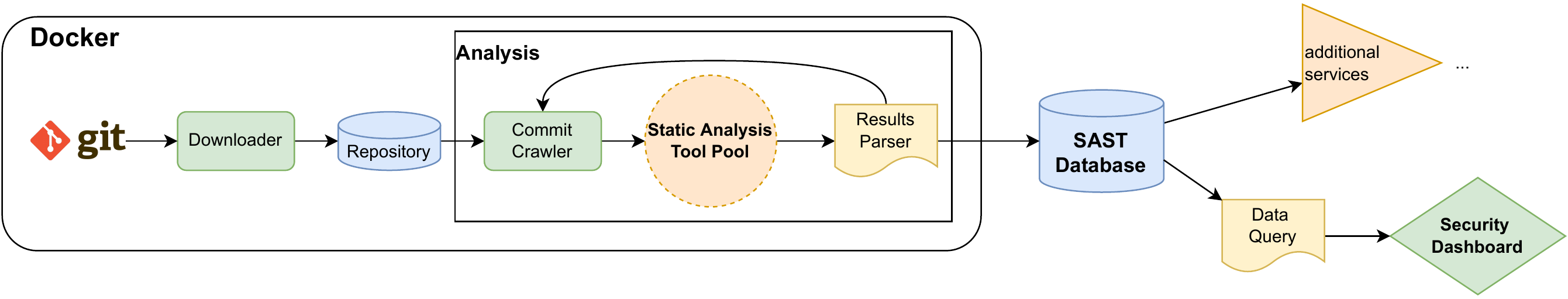}
	\caption{Platform overview.}
	\label{fig:platform-overview}
\end{figure*}

\section{SAST Platform Design}

The components of our analysis platform are mapped in \autoref{fig:platform-overview}. Starting from a Git URL, the script downloads the entire repository and extracts the repository state at a given commit. When the analysis is first run, the commits are downloaded and analyzed in chronological order. On all subsequent runs of the analysis, any new commits added since the last run are added to the analysis queue. The results are parsed and written to a database. In addition, the results can be displayed in our dashboard or retrieved from the database with other visualization tools such as Grafana \footnote{Grafana is a open source analytics and monitoring tool for databases. \url{https://grafana.com/}}.

\subsection{SAST Tools}

\begin{table*} %[h]
	\caption{Included static analysis tools.}
	\begin{center}
		\begin{tabular}{|l|l|l|l|}
			\hline
			\textbf{SAST tool} & \textbf{Category} & \textbf{Languages} & \textbf{Configuration} \\ 
			\hline
			Flowdroid & taint analysis & Android & \textit{java -jar soot.jar -s ./flowdroid/SourcesAndSinks.txt -o report.xml} \\
			Infer & formal verification & Java, Android, C, C++, iOS & \textit{infer run --results-dir \{resultDir\} --no-fail-on-issue -- \{compileConfig\}} \\
			MobSF & various & Android, iOS, Windows & \textit{mobsf --apikey \{Key\} upload \{Apk\} $>$ mobsf\_report.json} \\
			PMD & coding rules & Java, JavaScript & \textit{pmd -R rulesets/java/sunsecure.xml -failOnViolation false -f json}\\% \textit{-R rulesets/java/sunsecure.xml} \\
			Xanitizer & taint analysis & Java, Scala, JavaScript, TypeScript & \textit{xanitizer  generateDetailsInFindingsListReport=True overwriteConfigFile=True} % rootDirectory=\{repoDir\} findingsListReportOutputFile=\{resultFile\}
			\\ 
			\hline
		\end{tabular}
		\label{tab:sast-tools}
	\end{center}
\end{table*}

Our platform currently supports five SAST tools. \autoref{tab:sast-tools} shows the implemented tools, each with their category, the analyzed programming languages, and the predefined configuration. The categories range from coding rules to formal verification. As for the programming languages, in this work we mainly focus on the analysis of Android and Java. We selected these tools based on their suitability for analyzing the source code of the app.

We adapted the configuration of the tools (see \autoref{tab:sast-tools}) to issue the security-related alerts in order to detect vulnerabilities. The user of this tool can adapt it to project-specific conditions.
The generation of the best working configuration is still done manually by the developer, but with our platform, it is possible to test it across the entire development with a few simple steps.

The analysis itself runs in a Docker container to be platform-independent. To start the analysis, the script only needs the Git URL and the programming languages to be analyzed. We provide instructions to run the script as a systemd service to keep the database updated regularly.
Our scripts are written in Python, so they can be easily adapted for further data science applications.
We can include multiple repositories in the analysis queue.
The execution of the tools does not affect the output of the results in the dashboard, as both parts of the platform run independently. The analysis service ideally runs on a server, while the output takes place in the browser of a user client.

The tools we use have different requirements for the analyzed software. Either they use pure source code, compiled object code, or the analysis is directly integrated into the compilation process. Our platform currently supports Gradle and Maven builds by checking for the presence of the appropriate files in the main folder or sub-folders and executing the associated build command. After each commit, we perform a clean build to ensure all changes are applied correctly.
Additional SAST tools can be easily added by including the installation of the tool in the Dockerfile, defining the initialization of the analysis with the configuration, and writing a parsing method. We provide instructions for this in our repository.

We pay special attention to fault tolerance. If a SAST tool run fails or is aborted, the analysis is repeated on the next iteration. If the analysis repeatedly ends with an error code, the commit with that tool is skipped.

\subsection{SAST Database}

We use a relational database for storing the SAST warnings, repository, and tool information. The warnings from multiple repositories, including all tool information and their configurations, are stored in six related tables. This allows us to query and analyze the data through the backend of our dashboard. 
We use the commit hash or snapshot ID as the code state. The relational model can also be used as an interface to other data mining and analysis tools, as the authors have shown for the combined analysis of vulnerability and provenance data \cite{SchreiberProvenance2021}.

\begin{figure}[H]
	\centering
	\includegraphics[width=\linewidth]{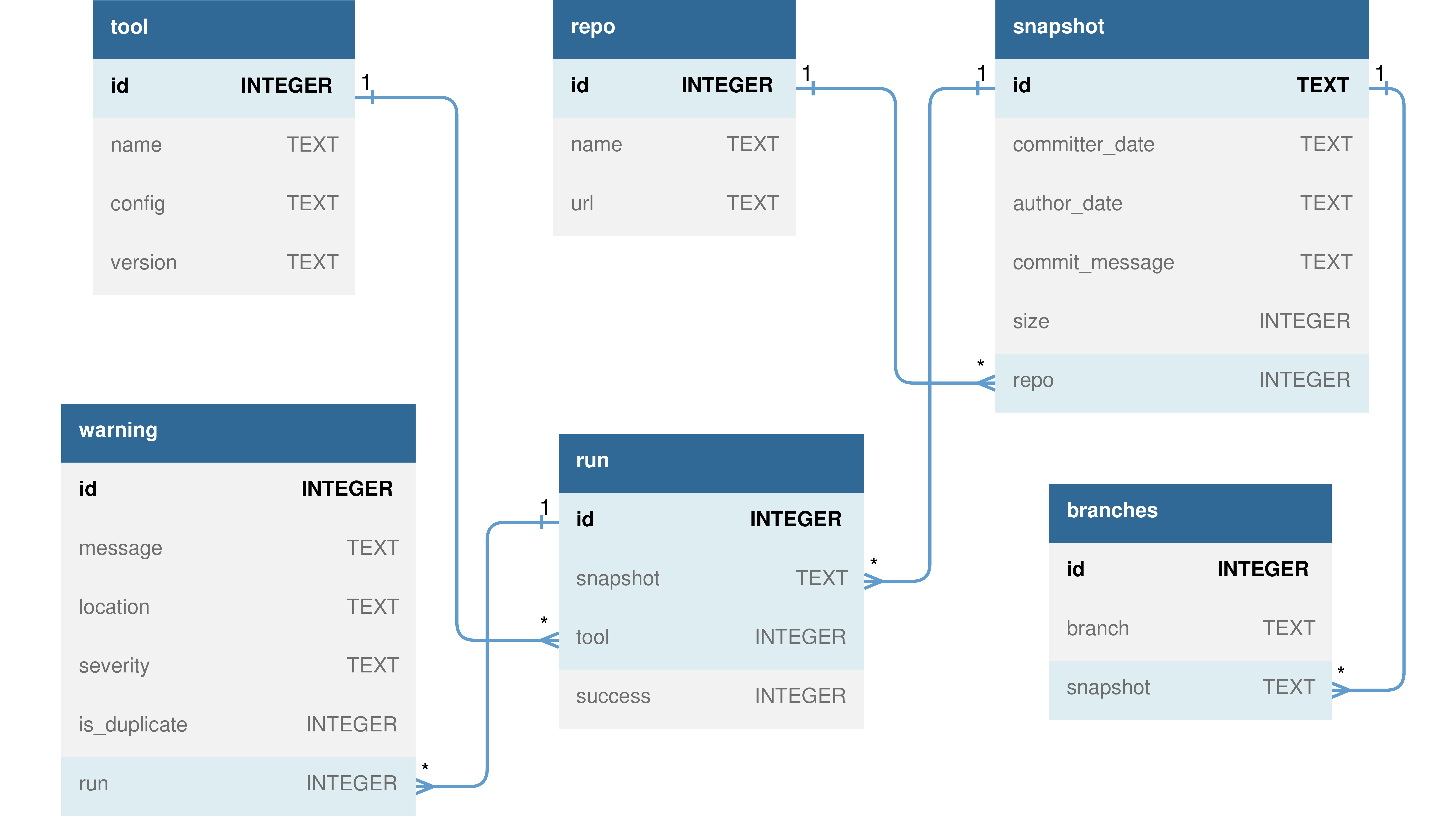}
	\caption{Relational schema of our SAST database.}
	\label{fig:database-schema}
\end{figure}

\autoref{fig:database-schema} shows the database schema. It includes the following tables:
\begin{itemize}[itemindent=-1em]
	\item[] \textbf{Repo} contains the commited repository name and its Git URL.
	\item[] \textbf{Snapshot} or commit contains the commit hash, commit date, author name, commit message, and code size measured by lines of code.
	\item[] \textbf{Branches} assigns a branch name to a specific \textit{snapshot}.
	\item[] \textbf{Tool} corresponds to a SAST tool, with a name, configuration and version number.
	\item[] \textbf{Run} corresponds to an execution of a \textit{tool} on a given \textit{snapshot} and confirms a \textit{success} flag which confirms a successful tool run or indicate errors \footnote{The user can read detailed error messages in the log files}.
	\item[] \textbf{Warning} is a collection of all SAST warnings encountered with a message, location as file path, severity (if available) and a duplicate check field.
\end{itemize}

The data types of the table fields are either text or integer.
We use 1-to-n relationships between the tables to be able to model the requirements.

After a tool run is complete, the results are returned depending on the tool, for example as a json file, and parsed appropriately so that they can be inserted with a SQL insert statement. 
The output of the tools, such as the warning message, location, and severity, varies by tool. It would be possible to normalize SAST warnings using, for example, the Static Analysis Results Interchange Format (SARIF) \cite{Anderson2018SARIF}. 
We do not normalize the values before writing them to the database, as we leave it open for future use.

The data model design we chose supports the visualization use cases we have for our dashboard. In the \textit{warning} table, we can easily filter the warnings, for example, by location or severity. The core table is \textit{run}, which has relationships with \textit{warning}, \textit{repo}, and \textit{snapshot}. An insight into the history of warnings can be obtained with a join of \textit{warning} and \textit{run.id} and a select of the \textit{snapshot.author\_date}.
In addition, it is possible to use the SAST data and correlate it with software metrics via the version history for deeper analysis, i.e., code churn correlated with the number of vulnerabilities \cite{shin2010evaluating}.
% We do not assume performance limitations due to the database design, since write operations to the database are always collected after a tool has completed an analysis for a commit. Furthermore, parallel write accesses are already prevented by SQlite.

\section{Dashboard}
We create a dashboard to visualize the security findings from the SAST database. \autoref{fig:dashboard-showcase} shows an example. The frontend should help the developers filter warnings more quickly or identify particularly buggy software modules. It is also intended to show a security trend, indicating whether SAST tools have been used in the past and whether the overall code quality has improved over the version history. The latter plays an important role, especially for legacy code.
% Currently, we focus on three basic types of diagrams:
Among others, the dashboard currently provides three basic visualizations:
\begin{itemize}[itemindent=-1em]
	\item[] \textbf{Version History Line Chart} This graph shows the number of SAST warnings in the version history to provide an overall security trend. This diagram is shown at the top left in \autoref{fig:dashboard-showcase}.
	\item[] \textbf{Alarm Type Tree Chart} This chart displays the most common alert types, either based on their warning message or a vulnerability type such as CWE (if present). This diagram is shown at the top right in \autoref{fig:dashboard-showcase}.
	\item[] \textbf{Hotspot Module Map} This mapping gives the viewer the ability to identify the hotspots, i.e., modules with a particularly high number of alerts.
\end{itemize}

Schreiber et al. provide a discussion of the visualization components in  \cite{SchreiberProvenance2021} and \cite{SchreiberDashboards2021}.
We visualize the results per tool because we cannot aggregate the values without normalizing the reports. Simple aggregation, such as counting warnings per code location, would be possible, but we cannot rule out the possibility that the report contains different bugs without parsing them. We also list the contents of the database to read the entire message at the bottom of \autoref{fig:dashboard-showcase}. We plan to integrate a direct link to the code location in an IDE.

\begin{figure} %[H]
	\centering
	\includegraphics[width=\linewidth]{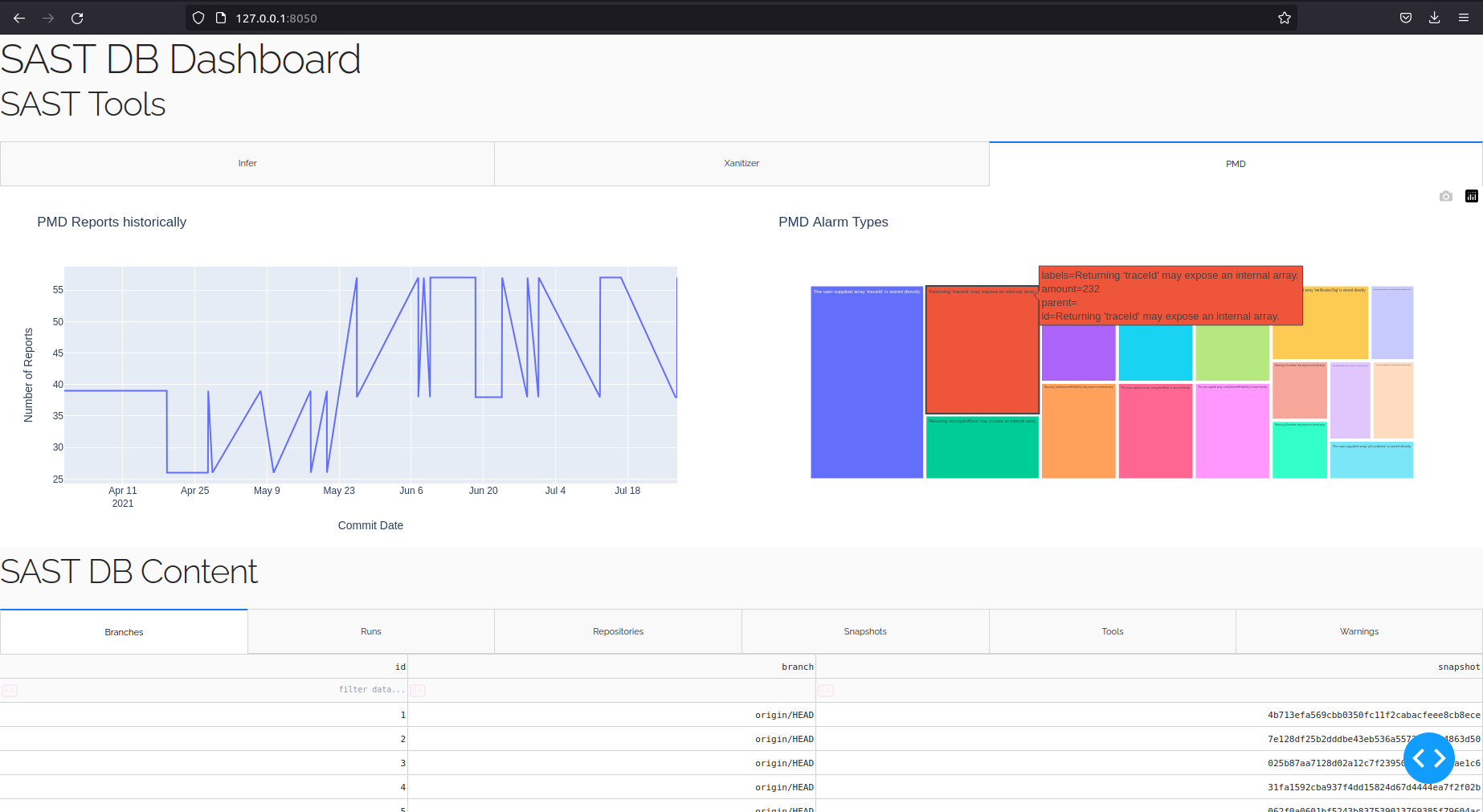}
	\caption{Dashboard showcase.}
	\label{fig:dashboard-showcase}
\end{figure}

\section{Limitations and Future Work}
Our analysis tool is ready-to-use for a huge number of Git-based projects. However, the platform currently focuses on Java projects and therefore supports only a subset of the respective SAST tools. In the future, we plan to extend this analysis platform to include more tools and other programming languages.

At the moment, we focus on collecting various warnings about the version history of a repository.
We do not perform a qualitative evaluation or normalization of SAST reports. Depending on which tools are used, there may be an increase in false positives but also true positives with multiple SAST tools.

Park et al. \cite{park2016battles} developed a method to reduce the number of false positives by including snapshots of execution environments. Such a method could also be included in our platform by extending it with dynamic analysis tools. In addition, a transformer network could help to prioritize the SAST warnings and thus provide filter functions for the developer \cite{sonnekalb2022deep}. 
Another next step is to plan and conduct a user study to help evaluate which elements in our dashboard are helpful to developers or can be improved. We would like to conduct this with focus groups to see which representations are most effective within a developer team. This method is very effective for improvement for UX design of dashboards  \cite{roberts2017give}. This also allows project-specific visualizations to be realized.

Static Analysis is used as a ground truth for machine learning applications \cite{russell2018automated}. By running our analysis platform on a large scale, it is possible to create a comprehensive dataset of SAST warnings. It can also be used to benchmark SAST tools.
In the future, we will use the SAST database to aggregate messages and additionally provide security and quality assessment by calculating code and repository metrics.

\section{Related Work}

Code review tools can be divided into SAST, DAST, IAST, and RASP tools. SAST tools are a form of white-box testing without running the code in a production environment. They are well suited for code checking during development or vulnerability scanning for quality assurance. In contrast, DAST (Dynamic Analysis Security Testing) tools correspond to black-box testing and can detect run time bugs by executing the analyzed software. IAST (Interactive Application Security Testing) tools combine SAST and DAST and perform code scans during runtime.
RASP (Runtime Application Self-Protection) is a technique with a runtime instrumentation technique that can block cyberattacks.
Our platform is solely based on SAST tools, but an extension with other tools is possible.

ReviewBot is a tool by Balachandran \cite{balachandran2013reducing}, that combines the results of multiple static analysis tools and automatically generates a code review. They collect user feedback for the reviews by conducting interviews and find that the majority of the automatically generated comments are useful. The tools are either linting or pattern-detection tools. We could not find an implementation to replicate this tool.

Nunes et al. \cite{Nunes2017} combine SAST tools for the application of web security, specifically SQLi and XSS vulnerabilities in WordPress plugins. They used a database of historic vulnerabilities for evaluation of the tool's performance. Their pipeline uses metrics to rank the tools and combine the outputs.
Our platform works with the current state of Git repositories, not historical data. Also, we are not limited to specific vulnerability types; we let the tools detect all of them. However, we also could not find their code for reproducibility.

Batyuk et al. \cite{Batyuk2011Using} build an analysis service for Android binaries that generate detailed reports. The binaries are downloaded from the Android Market, decompiled, and analyzed through data mining operations. Our analysis service starts with the source code, not the released version of an app, and therefore uses different SAST tools.

Scrub \cite{Holzmann2010Scrub}, short for Source Code Review User Browser, is a tool that combines classic peer code review with machine-generated analysis. It provides a user interface similar to our dashboard where developers can compare the SAST results with their manually reviewed code.
We did not plan to include manually reviewed code in our overview, as it is always biased by developer experience.

Meanwhile, Gitlab \cite{GitlabSAST} and GitHub \cite{GithubSAST} provide security scanning with SAST tools built into CI/CD. The advantage of such an analysis as ours is its high customizability and a database as output that can be used for more detailed and individual evaluation. It is also a novelty to be able to test different tool configurations with our platform and thus visualize security trends over time or over different software submodules.

Arai et al. \cite{arai2014gamified} use a gamification approach to motivate developers to use SAST tools and remove the warnings from the code. FindBugs' bug reports are classified into bug patterns, and a score is calculated per developer and per team. The score improves after corrections are made in the next iteration of the commit. This could be an interesting extension to our dashboard.

\section{Conclusion}
We provide an open-source static analysis platform as a Docker container that is easy to use and extensible with minimal effort. We also provide a dashboard that shows trends in version history. We published our code for reproducibility.

The novelty of our method is that it is based on the entire development history of the project, which makes it possible to derive security trends and identify hotspots. Tool configurations can be tested more easily by looking at how they have performed so far on the repository history.
Our analysis platform can help software developers in their daily work for the code review process and evaluate the version history of software projects for security and vulnerability analysis.
%, but also can also be used to build a comprehensive software analytics dataset.

%\newpage

\bibliographystyle{ieeetr}
\bibliography{sast_pipeline} 

\end{document}